# Supply Chain Digital Twin Framework Design: An Approach of Supply Chain Operations Reference Model and System of Systems

Jie Zhang[1], Alexandra Brintrup[1*], Anisoara Calinescu[2], Edward Kosasih[1], Angira Sharma[2]
1. Institute for Manufacturing, University of Cambridge
2. Department of Computer Science, University of Oxford

## Abstract:

Digital twin technology has been regarded as a beneficial approach in supply chain development. Different from traditional digital twin (temporal dynamic), supply chain digital twin is a spatio-temporal dynamic system. This paper explains what is 'twined' in supply chain digital twin and how to 'twin' them to handle the spatio-temporal dynamic issue. A supply chain digital twin framework is developed based on the theories of system of systems and supply chain operations reference model. This framework is universal and can be applied in various types of supply chain systems. We firstly decompose the supply chain system into unified standard blocks preparing for the adoption of digital twin. Next, the idea of supply chain operations reference model is adopted to digitise basic supply chain activities within each block and explain how to use existing information system. Then, individual sub-digital twin is established for each member in supply chain system. After that, we apply the concept of system of systems to integrate and coordinate sub-digital twin into supply chain digital twin from the views of supply chain business integration and information system integration. At last, one simple supply chain system is applied to illustrate the application of the proposed model.

## 1. Introduction

In the environment of Industry 4.0, the emergence and application of a large number of new technologies have made it possible to achieve more accurate and efficient supply chain (SC) planning, management and control. Recently, SC has been moved to a more electronic form with advanced information technologies. Many studies (Addo-Tenkorang and Helo, 2016; Moghaddam and Nof, 2018; Choi et al., 2018; de et al., 2020; Yadav et al., 2020; Yılmaz, 2020; Ivanov and Dolgui, 2020) have discussed emerging technologies under Industry 4.0 and their potential impact on SC, such as Internet of Things, smart & connected products, big data, artificial intelligence, additive technology, automatic robots, simulation technology, etc. The research results show that these emerging technologies have the advantages to be widely applied in the development of digital SC and can improve the performance of SC more efficiently, especially in risk control, optimisation, etc. So they have increased researchers' interest in trying to use these technologies to build supply chain control tower (SCCT) (Alias et al., 2014; Hofman, 2014; Liotine, 2019; Dalporto and Venn, 2020; Verma et al., 2020), supply chain cyber-physical systems (CPS) ( Klötzer and Pflaum, 2015a; Klötzer and Pflaum, 2015b; Frazzon et al., 2015; Tu et al, 2018; Klötzer, 2018; Hohmann and Posselt, 2019; Chen et al., 2020; Park et al., 2020), and supply chain digital twin (SCDT) (Srai et al., 2019; Ivanov et al., 2019; Wang et al., 2020; Marmolejo-Saucedo, 2020; Ivanov and Dolgui, 2020). These research focus on concept definition, framework design, model development and case analysis, etc. Challenges in production, transportation and inventory management have bene covered in different phases in SC management.

The SCCT is a physical or virtual dashboard that provides accurate, timely, and complete SC events and data. It operates the SC within and across organisations to coordinate all related activities. It also provides the end-to-end overall visibility of the SC and nearly real-time information to support decision-making (One Network, 2014). SCCT is a big data analysis platform and shared service centre. Both CPS and DT are effective means to achieve cyber-physical integration. CPS provides services such as real-time sensing, information feedback, and dynamic control through the integration and collaboration of computing, communication and control (3C) (Hu el at., 2012; Liu el at., 2017). Compared with DT, CPS emphasises the powerful computing and communication capabilities of the information world (Rajkumar el at., 2010), which can improve the accuracy and efficiency of the physical world. Through tight connection and feedback loop, the physics and computational processes are highly interdependent, which can ensure a reliable, secure, collaborative, robust and efficient manner to monitor physical entities (Lee, 2015; Liu el at., 2017). DT is another concept related to cyber-physical integration. DT is to create high-fidelity virtual models of physical objects in virtual space to simulate their behaviour in the real world and provide feedback (Grieves, 2014). DT reflects the two-way dynamic mapping process and provides a complete product digital footprint (Tao el at., 2018). Similar to DT, the feedback loop is very important in CPS. However, the CPS architecture focuses on control, and the information system, which may affect multiple physical objects (Zhu el at., 2011, Dillon el at., 2011; Hu el at., 2012; Monostori el at., 2016). DT focuses on the mirror model to provide a comprehensive physical and functional description of the component, product or system (Söderberg el at., 2017). The first and most important step is to create a high-fidelity virtual model to truly reproduce the geometry, physical properties, behaviour and rules of the physical world (Tao and Zhang, 2017). These virtual models are not only highly consistent with the physical parts in terms of geometry and structure, but also can simulate their temporal and spatial state, behaviour, and function (Debroy el at., 2017; Schleich el at., 2017). In addition, the model in the digital environment can directly optimise the operation and adjust the physical process through feedback (Vachálek el at., 2017). Using two-way dynamic mapping, physical entities and virtual models evolve together (Glaessgen and Stargel, 2012). Therefore, DT enables companies to predict and detect problems faster and more accurately, optimise processes, and get better solutions. Therefore, for the SC system, wherein many what-if scenarios are required for simulation and optimisation to reduce the uncertainty and fluctuation of each link, the DT system is a better choice.

However, the DT research related to SC mainly focuses on manufacturing (Tao el at., 2018; Zhou el at., 2020; Shao and Helu, 2020; Zhang el at., 2020; Lu el at., 2020) and logistics (Korth el at., 2018; Kuehn el at., 2018; Haße el at., 2020; Greif el at., 2020; Pan el at., 2020a; Pan el at., 2020b). Industrial Implementations related to SCDT are even fewer, while only some DT functions have been implemented. For example, DHL created one digital twin warehouse; Unilever built virtual versions of its physical factories with the power of machine learning and artificial intelligence (Emily, 2020; Sharma et al., 2020). For entire SC, studies are few. Srai et al. (2019) and Ivanov et al. (2019) discussed the concept of DT in SC. Srai et al. (2019) broadly defined the properties of SC, highlighting similarities and differences from the traditional factory perspective that places the emphasis on equipment and unit operations. Ivanov et al. (2019) analysed perspectives and future transformations to be expected in transition towards cyber-physical SCs, and demonstrated how digital technologies and smart operations can help integrate resilience of SC. Wang et al. (2020) discussed the benefits and potentials of DT used in SC, in comparison with the existing planning approaches, in terms of demand forecast, aggregate planning and inventory planning. Marmolejo-Saucedo (2020) developed a tool based on the concept of DT for decision-making in a SC. The objective is to share the information among the SC stakeholders in order to increase the visibility of products and processes. Ivanov and Dolgui (2020) created a generic structure of a digital SC twin for managing disruption risks. They combined model-based and data-driven approaches in the structure which allowed uncovering the interrelations of risk data, disruption modeling, and performance assessment. Overall, the current studies and applications are limited. The development of SCDT is still in its infancy as various challenges are yet to be mastered to put the DT meaningfully into practice. There is no general framework and specific feasible application methods for the entire SCDT system to explain what should be 'twined' and how to 'twin'.

SC is very complex and diverse. Different companies have different processes. The lack of standards has always been an obstacle during SC digital construction (Ivanov el at., 2017). The supply chain operations reference (SCOR) model is a complete framework with standardised and common definitions. It has been described as the "most promising model for SC strategic decision making (Huan et al., 2004). Therefore, this paper will combine SCOR model theory to establish a standardised SC module, design a general SCDT framework, and guide the construction of DT in the SC.

Another aspect of Industry 4.0 is that customer demand and business operation become more diversified. The corporate environment in the SC tends to be more complex, which puts forward higher requirements for the planning, management and control of each echelon in SC. It is necessary to develop an integrated SC to promote deep business integration and value chain reshaping in all echelons of the SC. In the actual SC system, in addition to the game of interests between enterprises, there are also differences in enterprise size and technical capabilities. Therefore, management and decision-making methods must be transformed from an individual model to a holistic model. When establishing SCDT, systematic approaches should be considered. Some studies have been conducted on the subject (Ivanov, Sokolov, and Kaeschel 2010; Ivanov 2018; Marmolejo-Saucedo, Hurtado-Hernandez, and Suarez-Valdes 2019). The system of systems (SoS) theory is one of the solutions to solve the problems in cooperation and integration (Choi et al., 2016), which will be applied in this study to develop the SCDT framework.

DT is a new technology and its construction cost is expensive. Not every member of the SC has the ability to support the construction of the system. However, the shortcomings of any echelon in the system will inevitably affect the performance of the entire system. Nowadays, enterprises have a certain level of information foundation and have their own diversified information systems. When building SCDT, how to make reasonable use of these existing systems is the key to the rapid implementation of SCDT. This is another problem that is to be solved in this paper. This paper will establish a general standard SCDT framework to guide how to 'twin' the SC, with the following main objectives:

1) Definition of twin content: digitise activities to cover spatio-temporal dynamic and agents' communication issues of SC in digital twin building.
2) Modularisation of the SC system: divide the SC into 5 standard blocks, by combining with the properties change caused by the flow of materials in the SC.
3) Block digitalisation: explain how to use SCOR to digitise basic activities in SC. It uses existing information system of SC members to make SCDT construction more economical and efficient.
4) Sub-DT construction for SC members: develop independent DT system for each SC member based on the obtained data-based virtual blocks.

5) SCDT integration and evaluation: build up an overall SCDT by integrating all sub-DTs in SC, both supply chain business integration and supply chain information system integration are proposed.
6) Application of SCDT framework: use a simple SC system to introduce detailed application procedure.

The rest of the paper is organised as: Section 2 introduces the relevant theoretical background. Section 3 designs the SCDT framework based on the theories of SoS and SCOR. Section 4 illustrates the application of the proposed framework into real SC case. Lastly, conclusions were summarised in Section 5.

## 2. Theoretical background

In order to build up the systematic framework, we introduce three theories focused on the design of SCDT framework, these are, digital twin, system of systems, and supply chain operations reference model.

2.1 Digital twin

The concept of DT was first given by Grieves in 2003 (Grieves, 2014). It refers to having a digital replica of real entities, e.g., people, process, physical assets, systems and devices (El, 2018). DT opened up a new way to synchronise physical activities with the virtual world. With the rapid development of Internet of things, DT has become a hot research topic. DT has been applied in a variety of industries recently, including product design (Tao el at., 2019), production line design (Zhang el at., 2017), digital twin workshop (Tao and Zhang, 2017), production process optimisation (Uhlemann el at., 2017), predictive maintenance and operation status management (Tao el at., 2018). It helps to improve the system performance, reduce cost, monitor the status and process, and predict the future condition using big data and machine learning techniques. General Electric (US), Siemens AG (Germany), Parametric Technology (US), Dassault Systèmes (France), Tesla (US), DHL (US), Unilever (UK), etc. also have applied DT to industrial practice (Schleich el at., 2017; Emily, 2020; Sharma et al., 2020). These companies use DT to increase their product performance, manufacturing flexibility and competitiveness.

Generally, there two types of focus when applying the concept of DT in research and industries: narrow and broad (Zheng et al., 2020). The distinction refers to the integration with the whole system.

Taking the application in manufacturing industry as an example, the narrow sense of DT focuses on visualizing the product only, that is, how to fully describe a potential or actual physical production dynamically. The features are data-driven, intelligent perception, virtual reality mapping (Tao et al. 2019) and cooperation interactive. The narrow sense of DT includes three major components, as shown in Figure 1, that are the physical space, the virtual space, and the information processing linking between them (Grieves, 2014). After the product data in physical space has been collected and transmitted to virtual model in virtual space, the received data information will be processed and send back to the physical space to achieve data mapping, which is usually real-time mapping.

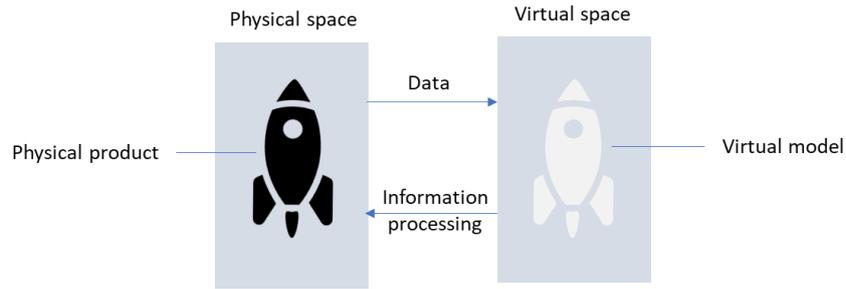

Figure 1 Narrow sense digital twin model

The broad sense of DT is an integrated system that can simulate, monitor, calculate, regulate, and control the system status and process, which is the functional extension of the narrow sense of DT. The features are individualisation, high efficiency and highly quasi-real. Scholars have summarised that this type of DT has five major components, that are: physical space, virtual space, connection, data, and service (Tao et al., 2018b). As shown in the structure illustration in Figure 2, the five components are equally important. The physical part is the basis for generating the virtual part. The virtual part is used to support the decision-making and management of the physical part through simulation. Data is at the centre of the entire DT, because data is the prerequisite for all functions. DT created new services to enhance the convenience, reliability, and productivity of an engineered system. Connection acts as a bridge to connect other parts, realising efficient interaction between different parts.

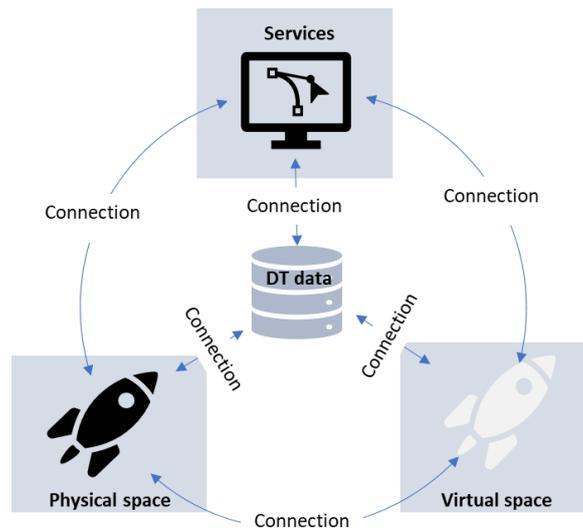

Figure 2 Broad sense digital twin model

2.2 System of systems

System of systems is 'a collection of task-oriented or dedicated systems that pool their resources and capabilities together to create a new, more complex system which offers more functionality and performance than simply the sum of the constituent systems' (Popper et al., 2014).

Although the research history of SoS theory is not long (Ge et al., 2014), it has become an important way to study complex and large-scale systems (Gorod et al., 2008). SoS can be described as: a large-scale system that combines multiple component systems with different functions to achieve a target objective or goal. Meanwhile, by combining these independent component systems, additional values can be brought to the overall system, like the formula '1+1>2'.

Scholars have summarised the feature of SoS as 'ABCDE' (Boardman and Sauser, 2008), which refers to Autonomy, Belonging, Connectivity, Diversity, and Emergence, respectively. Autonomy represents a system, as a component of SoS. It has the ability to make independent choices, including independence in action and decision. Belonging means that each independent component system has its own right and ability to choose as part of the entire SoS, based on its own needs, values, positioning and goals. Connectivity refers to the ability of each component system to be connected by other systems in the SoS. Diversity means that the overall SoS system has more diversified capabilities and system heterogeneity compared to fragmented individual systems. Emergence indicates that SoS is a dynamic system, and could generate new properties along with the interaction, development and evolution of the system.

Choi et al. (2016) verified that the SC is well-qualified as a SoS by applying the 'ABCDE criteria'. This is obvious because the philosophy of SoS is to form a comprehensive solution that maximises the goal by establishing interaction and collaboration between any system and individual. According to the overall plan, SoS can be applied to formulate each system and individual strategy. The SC itself is an integrated body. Within the SC, there are multiple echelons and multiple members. Members in each echelon are independent. However, in the SC system, it is necessary to enhance overall efficiency and reduce risks through cooperative behaviour. Eventually, the integration of the overall industrial chain based on core enterprises for logistics is formed.

2.3 Supply chain operations reference model

The SCOR model is a cross-industry standard supply chain reference model and SC diagnostic tool issued by the American Supply Chain Association. It provides comprehensive, accurate, and optimised standardised terms and procedures for SCs of various sizes and complexity. It consists of four parts: (1) The general definition of SC management processes, including plan, source, make and deliver. It is the basis for enterprises to establish SC performance and goals; (2) The performance index benchmarks corresponding to these processes; (3) The description of the "best practice" of the SC. It provides companies with the information that needed to successfully plan and determine the goals when improving the SC; (4) The selection of SC software product information and implementation of the configured specific SC (Council, 2017).

The SCOR model defines the SC as five major processes Plan, Source, Make, Deliver, and Return (Council, 2017). It has a classic process structure diagram, as shown in Figure 3. The process is divided into four levels, that are top level, configuration level, process element level and implementation level. Each process and the performance evaluation have been defined while the best practices of the SC and human resources plan have been also given in the model. The use of SCOR can enable the internal and external companies to use the same language to communicate SC issues, objectively evaluate its performance, and clarify the goals and directions for SC improvement.

Level 1: The top level includes five main processes, namely plan, source, make, deliver, return. Plan is the core of the SC and provides guidelines for SC operations. Source refers to the procurement of production or sales products. Make is the process from the production of raw materials to the completion of the finished product. Deliver represents the entire process from the factory to the customer to complete the purchase. Return is the process of returning goods from the customer or to the supplier.

Level 2: The configuration level focuses the process category on each main process from the top level. Taking the Deliver process as an example, it can be divided into 4 levels with 2 categories, namely: sD1: Make to stock (MTS), sD2: Make to order (MTO), sD3: Engineer to Order (ETO), sD4: Retail. These have standard definitions in the SCOR model.

Level 3: The process element level involves the specific processes. For example, the first process in MTS is 'sD1.1 Process Inquiry and Quote', and its description is to receive and respond to general customer inquiries and requests for quotes. SCOR shares 15 processes (from sD1.1 to sD1.15) to completely describe the entire process from customer order inquiry to the final billing to the customer.

Level 4: The implementation level is to perform activity. SCOR recommends that companies can define their own level 4 processes, which are generally specific activities in various industries and regions. For example, most companies need to perform an activity of 'receiving, entering and verifying customer orders', which is a Level 3 process (such as sD1.2). The Level 4 of the process is the step that describes how the company receives orders. It can be done via electronic data interchange (EDI), Fax, telephone or in a physical store. Each activity may require a separate Level 4 process description. Another step is to describe how the order is entered. EDI may automatically load data information, Fax and telephone orders are entered by the ordering platform, physical stores are processed by cashiers and so on. In short, there is no clear definition of the Level 4 process in SCOR, but the work is left to the company itself.

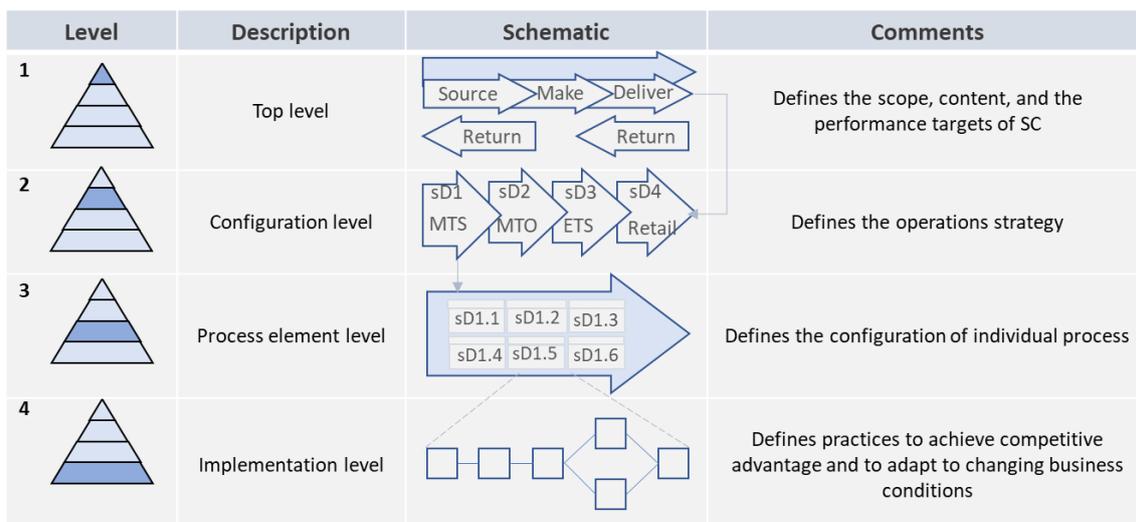

Figure 3 Process structure diagram of SCOR Model (Council, 2017)

The business processes are described based on how to satisfy a customer's demands. The model also provides a basis for how to improve those processes. It is one unique framework with standardized and common definitions, provides methodology and benchmarking tools, thus it can be adapted for simple or complex SCs across any industry. And it has been described as the "most promising model for SC strategic decision making (Huan et al., 2004). When SC is digitised, the lack of a unified standard is a key problem that has not been solved (Ivanov et al., 2017). This problem can be solved by introducing standardization processes and methods of SCOR into the design of the SCDT system.

## 3. Supply chain digital twin framework design

Supply chain is a network. Although it contains a series of physical entities, it is not an entity in a strict sense. The core of traditional DT is the physical entity, the twin is a replica of the physical entity, it is only temporal dynamic. While from the view of product's life, SC is one spatio-temporal dynamic network. Thus, SCDT is not the same as the traditional DT. Digitising the physical entities in SC is not sufficient for SCDT. The core things above these physical entities in SC are flows (physical flow, information flow and finance flow), which can describe the feature of spatio-temporal dynamic. The basis of these flows is a series of activities. These activities are associated with all physical entities in the SC, leading to the changes in their properties, and supporting the structure and dynamics of the SC. In SCDT, what we want to 'twin' is activities. Based on the digitisation of these activities, the planning and control of the SC system is realised through analysis, prediction, modelling and simulation methods.

In SC, activities are very diverse. There is no unified standard for the digitisation of the activities. This standard is crucial for building a general SCDT framework. It is important to know how to summarise and classify them according to the attributes of the activity (e.g., type, occurrence link). Therefore, the primary task before establishing SCDT framework is to analyse the components of the SC system. We need reasonably decompose the system into standard blocks with considering the activities. On this basis, the challenge of data exchange, individual and overall target design on building DT could also be addressed.

A real SC system includes multiple echelons. Each echelon may contain multiple members. Every member has his own demand of interests, which leads to diversified objectives in the entire SC system. Each member has a different scale and ability, and has varying importance in SC. How to rationally integrate the objectives among different members based on different abilities and also reflect the competition and cooperation among them have become the keys in the design of the SCDT framework. In addition, as of the complexity in SC system, how to use a unified and standard method to solve these diversified and non-uniform processes is the core in determine whether or not the framework can be practically applied. Finally, most companies have their own information systems. How to use these existing resources to reduce construction costs is also an important factor affecting the promotion of DT.

In order to solve these problems, we define the entire SCDT framework design procedure includes four parts: SC modularisation, basic activities digitisation, sub-digital twin (sub-DT) development and entire DT integration. SC modularisation part divides the SC system into standard business process modules. The basic activities digitalisation part solves the problem of how to adopt a unified and standardised method to digitise various basic SC activities into the proposed framework. It also explained how to use the existing information system in the construction of the new DT. The sub-DT development part is used to build the DT system corresponding to each member. The DT integration part is used to integrate the sub-DTs of multiple members to form the SCDT system. This could solve the imbalanced capabilities problem, which can promote the cooperation among all members.

## 3.1 Supply chain system modularisation

We divide the SC system into 4 levels from top to bottom, that are: SC system, member, module, and block, as shown in Figure 4. We use them to describe the changing of material/product's ownership and location properties.

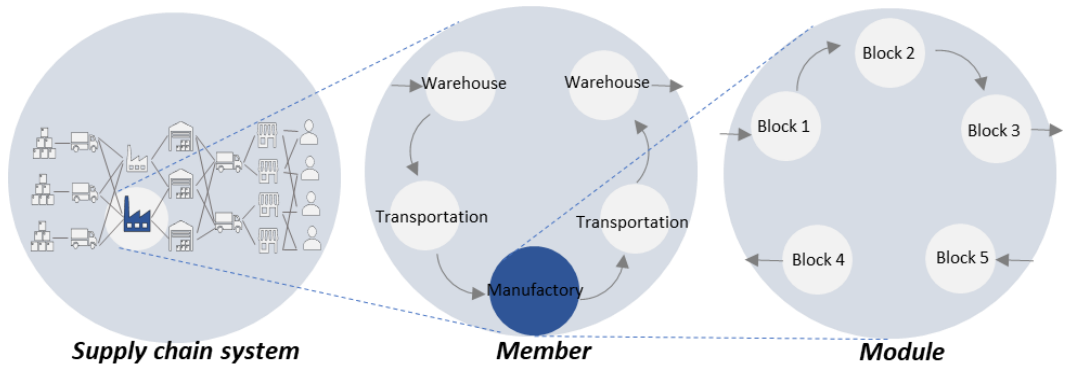

Figure 4 The 4-level structure of supply chain system

**Supply chain system**: A typical SC begins with the ecological, biological, and political regulation of natural resources, followed by the human extraction of raw material, and includes several production links (e.g., component construction, assembly, and merging) before moving on to several layers of storage facilities of ever-decreasing size and increasingly remote geographical locations, and finally reaching the consumer. Like shown in Figure 5, a SC system is a dynamic and complex network, which includes multiple echelons. Each echelon may contain multiple members.

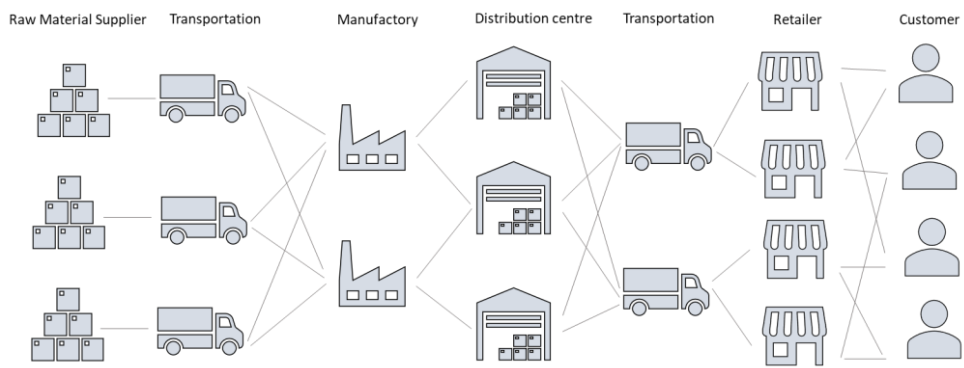

Figure 5 One structure of a supply chain system

**Member**: Member represents echelon operators in the SC, while they all have their own goals and considerations. SCs are diverse and the members in different SCs are different. In order to establish a unified standard framework that covers all types of SC, the idea of SCOR model is used to group the members in SC. In the SCOR model, the focused member has its supplier and customer, as shown in Figure 6, wherein the supplier and customer also have their own supplier and customer, respectively. Thus, a chain is formed naturally. Supplier and customer can be internal or external. In our framework, each member is independent on this design level, so that, there are no issues of internal and external problems. Meanwhile, the material/production flow transferring between two independent members needs to be completed by a transportation member. Therefore, a transportation member is added among the members of the original SCOR model.

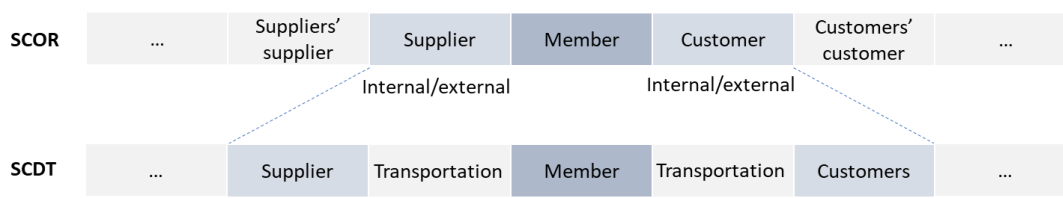

Figure 6 The members in SCOR and SCDT

**Module**: Module is the location carries of the material or product in the SC, which represents the geography changing. Each member can comprise one or more modules. For example, a complex factory member may contain these five modules, as shown in Figure 7. Each module is composed of multiple standard process blocks, as shown in Figure 8.

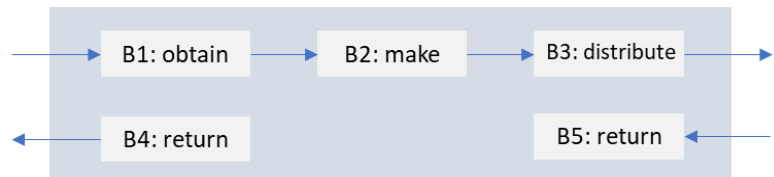

Figure 7 The structure of members

**Block:** Block is a process-set based on the business functions in each specific module. There are total five kinds of blocks in module according to the actual business, namely: B1: obtain (from upstream), B2: make, B3: distribute (to downstream), B4: return (to upstream), B5: return (from downstream). The specific number of blocks in each module is adjusted according to the actual business.

Figure 8 The structure of modules

3.2 Basic activities digitalisation

In SC system, the basic SC activities are various. To ensure the proposed standard framework can fit for the digitalisation of different SC activities, we refer the success experience in SCOR model. It is one unique standardized and common definitions framework with benchmarking tools. It is adapted for different SCs across any industry.

The SCOR model contains five main processes: plan, source make deliver and return, as shown in Figure 3. They correspond to the five blocks defined above, as shown in Figure 8. SCOR combines the actual business models of different companies to define the standard process elements of each process. For example, deliver includes a total of 15 sub-processes. For companies with more specific perform activities in sub-processes, SCOR recommends that companies should define them according to their own demand. In the proposed DT framework in this study, we utilise the existing information system to reduce the cost of DT construction and the difficulty in implementation. These existing information systems are just the embodiment of these diverse activities. We can directly apply these existing systems as a digital result of specific activities to form the underlying structure of SCDT.

We take the first sub-process element under the deliver process 'digitalisation' as an example. Assuming the business mode is 'Make to Stock', this sub-process is 'sD1.1 Process Inquiry and Quote'. Checking the SCOR model, we can get all the information about the sD1.1 process, including 'Practices' and 'Metrics ', as shown in Figure 9. The best practices related to the sD1.1 process are BP.114 and BP.176. According to the previous explanation (relationship between SCOR process and proposed blocks), sD1.1 is the process under 'Obtain' block. The practices include the details activities, which is what we need to digitise. Thus, we need to establish practices BP.114 and BP.176 in DT. For companies, these practices are existing information systems. During the construction of DT, the business improvement can refer to the specific provisions of SCOR on BP.114 and BP.176; data quality requirements can be applied to higher-end data collection internet of things devices.

SCOR model also includes a set of bottom-up evaluation system. In the process of sD1.1, SCOR designs two indicators to measure the performance, namely RS.3.100 and CO.3.14 under 'Metrics'. The description of RS.3.100 is 'Process Inquiry & Quote Cycle Time'. It is the average time for processing inquiries and quotes. The description of CO.3.14 (Order Management Costs) is the cost of order management. After the digital mapping based on 'Practices', we can obtain these two types of time (RS.3.100) and cost (CO.3.14) indicators to evaluate the process of sD1.1 in customer order inquiry and quotation. These indicators will be fed to support the monitoring and decision-making functions in virtual libraries, which is used in the establishment of sub-DT and modules.

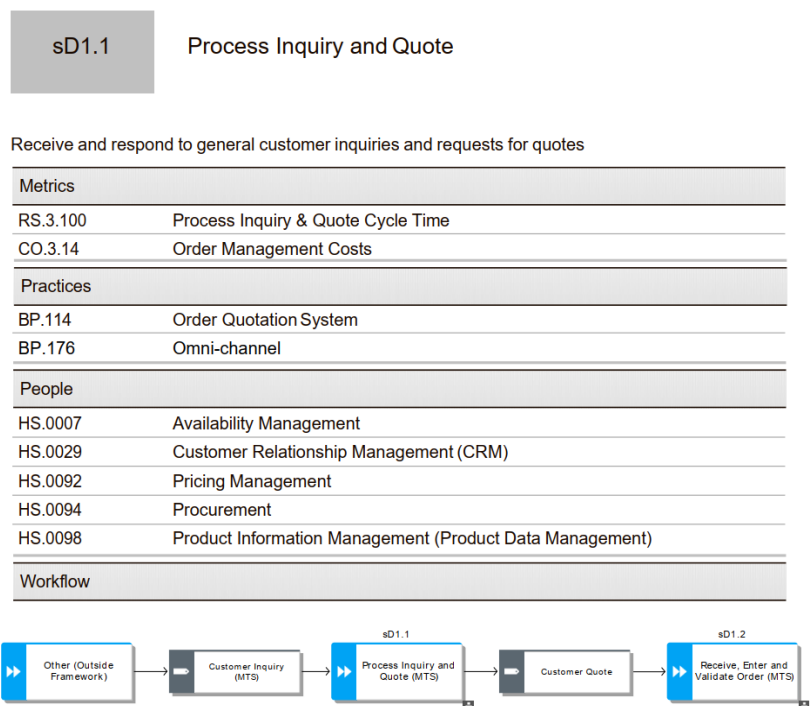

Figure 9 The detail information of sD1.1 (Council,2017)

### 3.3 Sub-Digital Twin development

Sub-DT is the DT of each member in the SC system. As each member is an independent individual, the sub-DT can be regarded as an independent DT. The framework is built from the bottom to up. Starting from the smallest standard unit block level, the data mapping of all blocks is realised through DT technology to form a narrow sense of DT virtual system; then according to the business rules, a virtual model is established by compromising multiple blocks systems to form a module. Lastly, an independent sub-DT framework is developed by combining relevant modules. Similar to traditional DT systems, the framework includes three parts: physical space (strictly speaking, it is not physical, but assumed to be generalised physical), virtual space, and information-processing layer, which is shown in Figure 10.

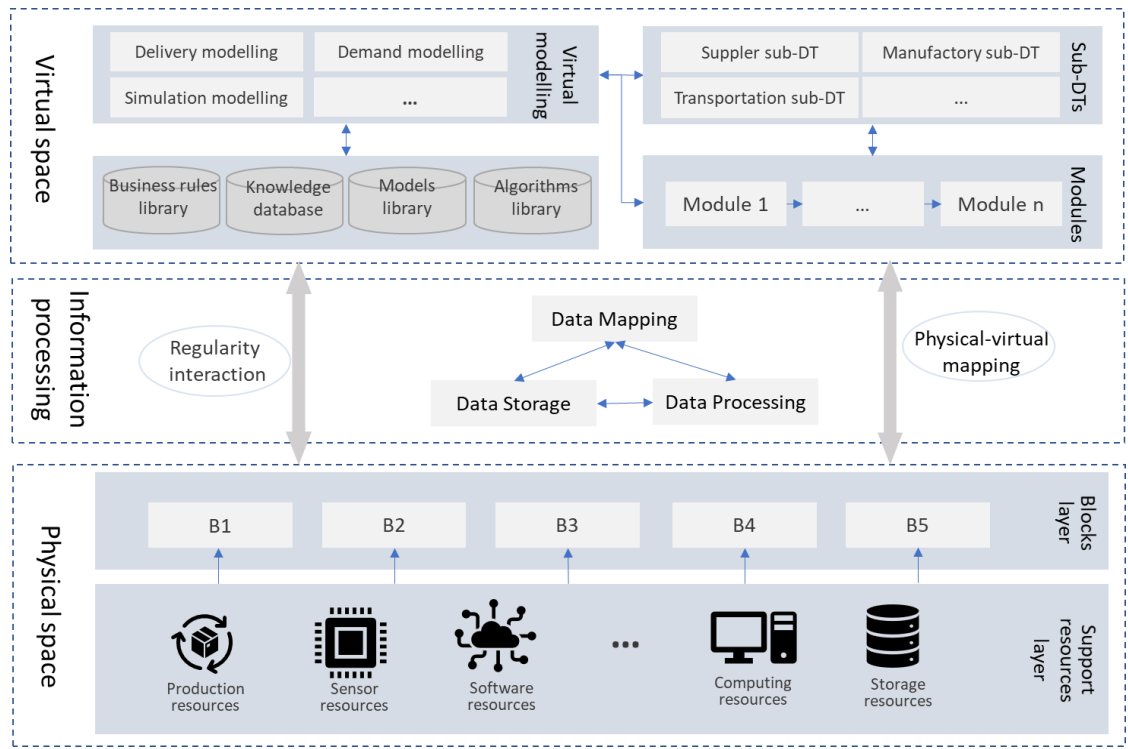

Figure 10 The framework of Sub-DT

**Physical space**

The physical space of SC is a complex and diverse environment, which includes members, processes, materials and rules, etc. For simplicity and clarity, Figure 10 uses blocks as an example. The detailed information in blocks has been introduced above. Support resource layer includes all kinds of objects that support the digitalisation in blocks, such as production resources, computing resources, software resources. All of them are connected by IoT technology to collect and integrate the data of physical world.

**Information processing**

The information processing layer is the channel connecting physical space and virtual space, the bidirectional mapping and interoperation of physical space and virtual space are realized through the data interaction in this layer. There are three main function of this layer: data mapping, data storage and data processing. Data mapping supports the synchronous mapping of physical data and virtual SC blocks. We need to store the data both from physical space and virtual space. Data processing includes data pre-processing, data analysis and mining. Because of the diversification of data collection sources and technologies, raw data requires pre-processing and noise reduction. The huge amount of data requires proper data analysis and mining approach (e.g., from feature level and decision level) to improve the quality and dimensions.

**Virtual space**

In addition to the data mapped from the physical space, the virtual space includes virtual modelling environment and the DT system. There are interactions between them. Virtual modelling environment provides various virtual models for modules and sub-DTs, such as production models, transportation models, and predictive models. These models are established through a series of libraries and block mapping data, be stored in the database by using corresponding interfaces. Combining various models, methods, and mapping data, the modules and sub-DTs are obtained. The DTs, as the real mapping of physical entities, can not only realise the visualization of agents, but also realise the simulation of complex systems. When conflicts and disturbances occur in physical space, virtual models can be tested in real time or even predict them and feed the information back to the physical space.

### 3.4 Digital Twin integration

According to the above design of sub-DT, each sub-DT has the only one operating entity. Within each sub-DT, there are no issues on data sharing between different operators and inconsistent goals. This can be applied to each member in SC to develop the relevant sub-DTs. The entire SCDT system structure is in a 3D format, as shown in Figure 11. Each layer is a sub-DT system, and the flow of SC connects the physical space of each member in an orderly series. The sub-DTs are different, as the business, scale and even abilities of each member are different. We need use SoS theory to let them collaborate based on vertical integration and horizontal coordination. The SC system of systems integration includes supply chain business integration and supply chain information system integration.

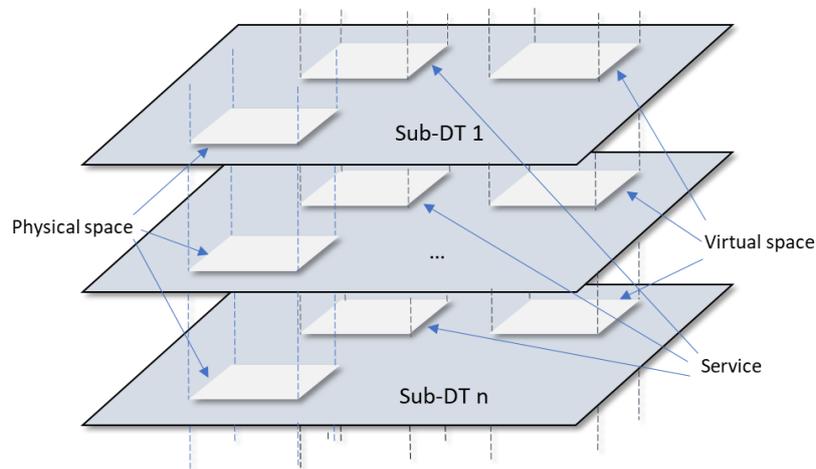

Figure 11 The SCDT structure

#### 3.4.1 Supply Chain Business Integration

The integration in the layer of business mainly relates to the SC operation modes and goals, includes the cooperation style, profit division way, enter & exit mechanism and so on. There are already lots of approaches and models on them. In order to let the business integration at SoS level by using these methods, some design principles are explored firstly. Observe that some of these principles are adapted and refined from the related literature (Maier,1998; Phillis et al., 2010; Choi, and Shen, 2016).

Principle 1. Setting goals and identifying contributions: Define the goals of the whole SC and identify the functions and contributions to the goals by members.

Principle 2. Adopting the policy triage: A SC has many members. There might exist many underperformers or the so called 'weak links'. A SoS should eliminate the weak links, keep and develop the needy.

Principle 3. Achieving coordination and integration: each member of SC is managed by an independent member. It is a fact that these members may not be integrated together. A SoS should ensure cooperation among its element systems.

Principle 4. Managing risk to establish stability: Make use of technologies to alarm the possibility of the collapse of the whole SC system of systems.

Principle 5. Incorporating crowdsourcing: Aim at collecting inputs and data from non-expert contributors to enhance knowledge.

In addition, based on the proposed principles, we combine it with the 'ABCDE' features of SoS and create the 'supply chain system of systems matrix', which is depicted in Table 1. It is the tool to estimate whether some specific issues and actions are satisfied with SoS requirements and provides practical guidance to practitioners.

Table 1 Supply chain system of systems matrix

|   | **Principle 1** | **Principle 2** | **Principle 3** | **Principle 4** | **Principle 5** |
|---|---|---|---|---|---|
| **A** | Goal alignment | Let members freely form partnership | Provide incentives measures | Provide advice and support to each member | Members work together |
| **B** | Set common goal | Keep strong links, help the needy | Provide support to develop integration and collaboration | Set mutually beneficial risk management schemes | Take collaborative action |
| **C** | Get connected on goals & functions | Communicate to ensure all links are healthy | Develop reliable connectivity | Tight communication | Connectivity is important |
| **D** | Treasure member's contribution | Allow individual diverse functions | Consider diversity issue in setting coordinated decisions | Provide a risk scheme for all members | different levels of participation |
| **E** | Goal adjustment | Regular checking and adjustment | Set robust coordination schemes | Identify threats from changes and responds | Adjust schemes dynamically |

3.4.2 Supply Chain Information System Integration

There are many information system integration patterns those proposed by previous researchers can be used to support the SoS-SCDT integration. But the choice of pattern is greatly related to the integration form. For example, how data will be shared and how control will be managed. Data may be shared or isolated. Control may be strictly hierarchical or no one system controls another. Such considerations lead to the results in Table 2.

Table 2 'Data' and 'Control' in system of systems

|   |   | **Data** | |
|---|---|---|---|
|   |   | Shared | Isolated |
| **Control** | Hierarchical | Information System with Shared Memory | Traditional Information System |
|   | Autonomous | Data-Centric System | Agent-Based System |

We hope the final SoS-DT reaches the optimal extremes dimension, but most real-world SoS will be somewhere in between. There are many features those determine the integration form, such as integration level, information exchange type, abstraction integration way, interaction style. Two of them are described as example below.

1) **Integration level**

Data level: data exchange or the common data access exists in entire system. one system can use the information from another system as part of its normal processing.

Service level: one system makes use of the capabilities of another one.

Business process level: there is a complex interaction among the different systems.

**2) Information exchange type**

Inform: one system provides information to one or more systems. It is one-directional information exchange.

Sync: two (or more) systems exchange information to keep each other in sync which is bi-directional information exchange.

Control: one system is predominant who can determine how to act based on the information from others.

Negotiation: Multiple systems negotiate to decide how information exchange to achieve their particular purpose.

After understanding the features, we can determine the integration pattern based on the real business. Some of common patterns are summarised as shown in Table 3.

Table 3 Summary of common integration patterns

| Pattern | Integration Level | Information Exchange Type | Defined by |
|---|---|---|---|
| Service-Oriented Architecture | Data/ Service/ Business process | Inform/ Sync | Bass et al., 2003 |
| Publish-Subscribe | Data/ Service/ Business process | Inform/ Sync/ Control/ Negotiation | Bass et al., 2003 |
| Canonical Data Model | Data | Inform/ Sync | Hohpe and Woolf, 2004 |
| Dynamic Router Pattern | Data/ Service | Inform/ Sync | Hohpe and Woolf, 2004 |
| Blackboard | Business process | Inform/ Sync | Buschmann et al., 2007 |
| Data Warehouse | Data | Inform/ Sync | Köppen et al., 2011 |
| Collaborative Virtual Environments | Business process | Sync/ Control | Churchill et al., 2012 |
| Remote Facade | Service | Inform/ Sync | Fowler, 2012 |
| Remote Process Invocation | Service | Sync | Kazman et al., 2013 |
| Batch Data Synchronization | Data | Inform/ Sync | Kazman et al., 2013 |

Finally, the integration quality of information system also should be ensured. Below we identify some important qualities in integration patterns those can be used as evaluation indexes of the integration, as shown in Table 4.

Table 4 Integration quality evaluation indexes

| Index | Description |
|---|---|
| Reliability | communication should be informed if the integration breaks down |
| Performance | perform adequately, do not require too many intermediate steps and data transfers |
| Security | must assure the source of the data, no alteration of the exchanged information may occur |
| Availability | the integration source / destination remains available |
| Scalability | the integration is scalable across large numbers of systems |
| Manageability | Be easy to manage the integration |
| Consistency | ensures the validity and integrity of the data shared |

## 4. Case study

This section is to illustrate how to apply the proposed SCDT framework into SC system by a simple case study. We consider a three-echelon simple SC with six members: 2 suppliers, 1 transport member, 1 manufactory and 2 retailers. Suppliers provide different type of raw materials. The raw materials from supplier 1 are shipped by one third party transport member, while the other delivery services among members are provided by manufactory. Manufactory stores raw materials and processes then into products, then stores them for purchase by retailers. Retailers are responsible for sales after purchasing the products.

### 4.1 Supply chain modularisation

Based on the division method of section 3.1 and the business functions, the entire SC can be divided into a structure composed of standard blocks as shown in Figure 12. Each supplier, transport member and retailer contain one module; manufactory includes three modules.

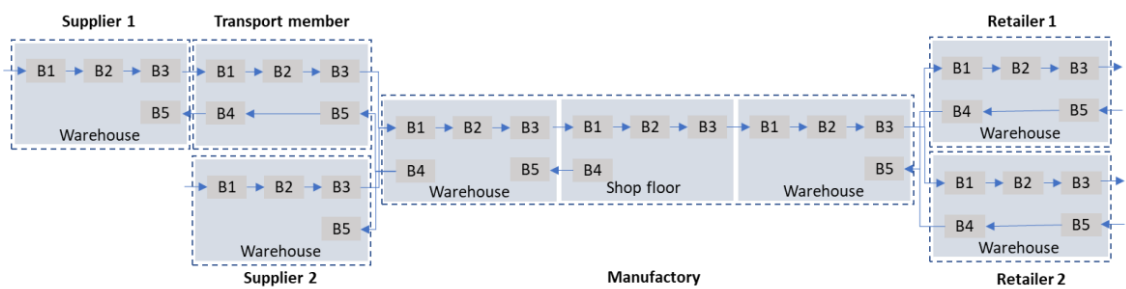

Figure 12 The block structure of the entire supply chain

### 4.2 Virtual block construction

Using the shop floor module as an example. According to the definition of the processes in the SCOR model, the processes can be further refined to get Figure 13. The virtual block contains physical model and dynamic model. The physical model describes the shape and size of the machine which is related to the real shop floor. Dynamic model defines all detail activities in shop floor.

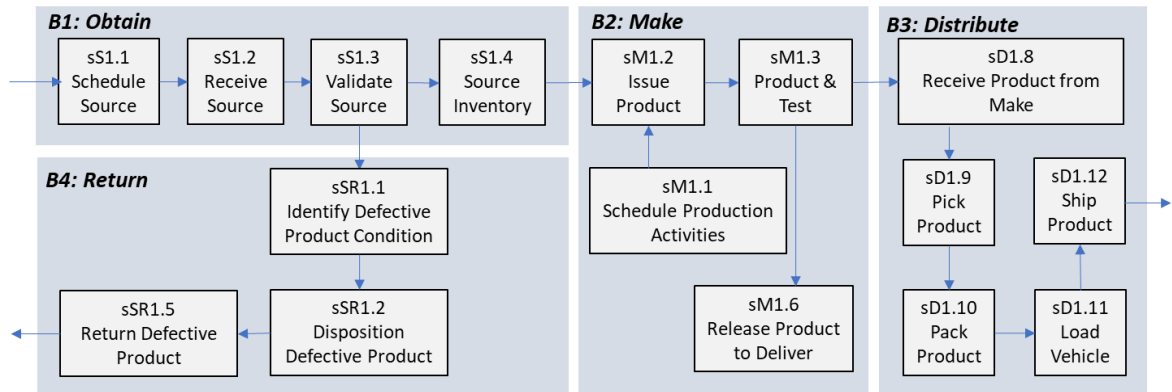

Figure 13 The processes of shop floor block

For B2 (Make), checking the best practices and evaluation indicators of these processes from SCOR model, we can extract Table 5. SCOR model proposes best practice for each activity, so these practices can fully describe the dynamics on the production line.

After mapping the physical machine, the practices will be applied to digitise the dynamics. They are software systems and tools those are used by manufactory or should be applied to reach the best performance. The software systems can be directly integrated into virtual block construction. For example, the 'Automated Data Capture' is one interface of sub-DT system, 'Production Line Scheduling' is one component of virtual library and database as shown in Figure 10.

Table 5 The best practices and metrics of block 2 (Make)

| Process | Practices | Metrics |
|---|---|---|
| sM1.1 | Production Line Scheduling | Schedule Production Activities Cycle Time<br>Capacity Utilisation |
| sM1.2 | Kanban<br>Production Line Scheduling<br>Lot Tracking<br>Automated Data Capture<br>Mixed Mode/Reverse Material Issue | Issue Material Cycle Time<br>Packaging as % of total material |
| sM1.3 | Lot Tracking<br>Automated Data Capture | Fill Rate<br>Warranty and Returns<br>Produce and Test Cycle Time<br>% of production materials reused<br>Capacity Utilisation |
| sM1.6 | Perfect Pick Put away | Release Finished Product to Deliver Cycle Time |
| sM1 | Single-Minute Exchange of Die<br>Business Rule Review<br>MTO Order Fulfilment Strategy<br>Mobile Access of Information<br>Bar coding/RFID | Make Cycle Time<br>Current manufacturing order cycle time<br>Upside Make Adaptability<br>Downside Make Adaptability<br>% of labor used in manufacturing<br>Cost to Make<br>Direct Material Cost<br>Indirect Cost Related to Production<br>Direct Labor Cost<br>Risk Mitigation Costs |

### 4.3 Sub-digital twin construction

Sub-DT includes physical space, virtual space and information-processing layer. Information-processing layer stores the virtual blocks data and guarantees the real-time mapping and interaction between physical and virtual space. Data processing methods are applied in this layer to make the communication more efficiency.

Virtual blocks above are combined to form the data mapping of modules and sub-DT. Each practice has its performance indicator as shown in Metrics column (Table 2). These quantitative indicators and the models behind them are stored in library to build the virtual modelling. A series of artificial intelligence methods are also necessary to be added in library to analyse and predict patterns in SC.

Based on the actual business of the SC member, we integrate data mapping system, virtual modelling and the indicators to build sub-DT system. It can analyse historical data, describe current situation and predict future trend. These abilities are used to simulate what-if scenarios and feedback to system, then help SC member to monitor and make optimal decision in business.

### 4.4 Supply chain digital twin construction

DT integration is the last step. Following the explains in section 3.4, we first should be clear of the goal and optimal operation strategies for the entire SC. The choice of each member's strategies can be designed based on the detail information of SC. Meanwhile, the SoS matrix will be used to evaluate whether the strategy is suitable. Then the SCDT integration with SoS will achieved at the layer of business.

The strategies with considering SoS will not only be transferred to models and rules to combine the information systems, but also support the decision of information system integration pattern. For example, if SC business SoS integration result shows that there is a complex interaction among the different members and all members exchange information to keep each other in sync. Then checking with Table 3, we can find 'Blackboard', 'Service-Oriented Architecture' or 'Publish-Subscribe' can be selected as the information integration pattern. It is sure that some other patterns those are not listed in Table 3 also may be a good choice according to real SC business. At last, make sure the integration quality indexes are met when apply the selected pattern. In this way, the SCDT with SoS is built.

### 5. Conclusion

This paper addressed the topic of DT development in SC. Different from traditional DT, SCDT is one spatio-temporal system. The digitalisation of physical entities themselves is not enough to describe the feature of SC. We proposed to 'twin' activities to solve this issue and show how to 'twin' them. The research gap of no universal SCDT framework has been filled. We introduced the relevant theories that need to be used when building the framework. To design one standard method to digitise SC system, the entire system has been firstly decomposed into standard blocks. We divided the entire system into 4 levels from top to bottom, that are: SC system, member, module, and block. The rules of setting these 4 levels are based on the product's properties changing over the time. To solve the challenges of diversified process digitalisation, we use the context from SCOR model to digitise basic SC activities and extract evaluation indicators within each block. Meanwhile, it is explained that the existing information system of SC members can be applied to the construction of DT, thereby reducing construction cost and implementation difficulty. Based on the virtualized blocks, we establish the standard individual DT for each member in SC system which includes physical space, virtual space, and information-processing layer. The theory of SoS is applied to integrate different sub-DTs into SCDT from the views of SC business and SC information system. At last, we used a simple SC example to illustrate how to apply the proposed framework in real SC.